# Achiral symmetry breaking and positive Gaussian modulus lead to scalloped colloidal membranes

Thomas Gibaud[1,2], C. Nadir Kaplan[1,3], Prerna Sharma[1,4], Andrew Ward[1,5], Mark J. Zakhary[1], Rudolf Oldenbourg[6,9], Robert B. Meyer[1], Randall D. Kamien[7], Thomas R. Powers[8,9] and Zvonimir Dogic[1]

[1]*The Martin Fisher School of Physics, Brandeis University, 415 South Street, Waltham, Massachusetts 02454, USA*
[2]*Univ Lyon, Ens de Lyon, Univ Claude Bernard, CNRS, Laboratoire de Physique, F-69342 Lyon, France*
[3]*John A. Paulson School of Engineering and Applied Sciences, Harvard University, Cambridge, MA 02138, USA*
[4]*Department of Physics, Indian Institute of Science, Bangalore 560012, India*
[5]*Program in Cellular and Molecular Medicine, Boston Children's Hospital, Boston, MA, USA*
[6]*Marine Biological Laboratory, 7 MBL Street, Woods Hole, MA 02543, USA*
[7]*Department of Physics and Astronomy, University of Pennsylvania, Philadelphia PA 19104, United States*
[8]*School of Engineering, Brown University, Providence, Rhode Island 02912, USA*
[9]*Department of Physics, Brown University, Providence, Rhode Island 02912, USA*

**Abstract:** In the presence of a non-adsorbing polymer, monodisperse rod-like particles assemble into colloidal membranes, which are one rod-length thick liquid-like monolayers of aligned rods. Unlike 3D edgeless bilayer vesicles, colloidal monolayer membranes form open structures with an exposed edge, thus presenting an opportunity to study physics of thin elastic sheets. Membranes assembled from single-component chiral rods form flat disks with uniform edge twist. In comparison, membranes comprised of mixture of rods with opposite chiralities can have the edge twist of either handedness. In this limit disk-shaped membranes become unstable, instead forming structures with scalloped edges, where two adjacent lobes with opposite handedness are separated by a cusp-shaped point defect. Such membranes adopt a 3D configuration, with cusp defects alternatively located above and below the membrane plane. In the achiral regime the cusp defects have repulsive interactions, but away from this limit we measure effective long-ranged attractive binding. A phenomenological model shows that the increase in the edge energy of scalloped membranes is compensated by concomitant decrease in the deformation energy due to Gaussian curvature associated with scalloped edges, demonstrating that colloidal membranes have positive Gaussian modulus. A simple excluded volume argument predicts the sign and magnitude of the Gaussian curvature modulus that is in agreement with experimental measurements. Our results provide insight into how the interplay between membrane elasticity, geometrical frustration and achiral symmetry breaking can be used to fold colloidal membranes into 3D shapes.

**Significance Statement:** A number of essential processes in biology and materials science, such as vesicle fusion and fission as well as pore formation, change the membrane topology and require formation of saddle-splay surfaces. The energetic cost associated with such deformations is described by the Gaussian curvature modulus. However, since conventional lipid vesicles almost always appear as closed structures of constant topology, experimental measurements of the Gaussian curvature modulus are challenging. We show that flat 2D colloidal membranes comprised of achiral rods are unstable and spontaneously form scalloped edges. Quantitative analysis of such instability estimates the Gaussian curvature modulus of colloidal membranes. The measured sign and magnitude of the modulus can be explained by a simple entropic excluded volume argument.

**Introduction:** The possible configurations and shapes of 2D fluid membranes can be described by a continuum energy expression that accounts for the membrane's out-of-plane deformations as well as the line tension associated with the membrane's exposed edge (1, 2). Since an arbitrary deformation of a thin layer can have either mean and/or Gaussian curvature the full theoretical description of membranes, in principle, requires two parameters, the bending and Gaussian curvature moduli. However, lipid bilayers almost always appear as edgeless 3D vesicles, which further simplify theoretical modeling. In particular, integrating Gaussian curvature over any simply closed surface yields a constant (3). Thus, the shape fluctuations of a closed vesicle only depend on the membrane bending modulus. Consequently, experiments that interrogated mechanics or shape fluctuations of vesicles provided extensive information about the membrane curvature modulus and how it depends on the structure of the constituent particles (4-6). In comparison, significantly less is known about the Gaussian modulus, despite the significant role it plays in fundamental biological and technological processes such as pore formation as well as vesicle fusion and fission (7-11).

Recent experiments have demonstrated that in the presence of a depleting agent monodisperse rods robustly assemble into one-rod length thick 2D membranes, with in-plane liquid order (12-16). Although more than two orders of magnitude thicker than lipid membranes, the deformations of both colloidal monolayers and lipid bilayers are described by the same elastic energy. However, in contrast to conventional membranes that fold into 3D vesicles, colloidal membranes appear as open structures. This presents a unique opportunity to explore the elasticity of 2D fluid sheets, a geometry for which both the Gaussian modulus and edge energy play an important role. It might seem that a flat circular shape would always minimize the free energy of colloidal membranes. This is not necessarily the case. We explore the possible shapes of colloidal membranes and demonstrate an unexpected connection between the membrane's edge structure, Gaussian curvature, and the chirality of the constituent rods.

The boundary conditions of colloidal membranes require twisting of the rods at the edge, and this twist penetrates into the membrane interior over a characteristic length scale (16-18). For membranes composed of single component chiral rods, the handedness of the edge twist along the entire circumference is uniform and dictated by the microscopic chirality of the constituent rods. With decreasing chirality, which is accomplished by mixing rods of opposite handedness, flat 2D circular membranes become unstable, and instead develop complex scalloped edges. In this limit edge-bound rods exhibit achiral symmetry breaking, forming domains of opposite twist that are separated by cusp-like point defects, where the membrane escapes into the third dimension. The exact structure of the scalloped edge is determined by the competition between the line tension and the Gaussian curvature modulus. Line tension favors circular flat membrane that minimizes the exposed edge. In comparison, an undulating scalloped edge create excess Gaussian curvature and is thus favored by membranes that have positive Gaussian moduli. Thus, observations of scalloped edges demonstrate that Gaussian modulus of colloidal membranes is positive. Tuning the membrane's chiral composition effectively controls the interactions between cusp defects that can be either attractive or repulsive. Measurements of these interactions leads to an estimate of the Gaussian curvature modulus that is in agreement with the predictions of a simple theoretical model .

## 1. Structure of colloidal membranes

Our experimental model system is colloidal membranes that spontaneously assemble in a mixture of dilute monodisperse rod-like viruses and non-adsorbing polymer Dextran. The viruses alone interact through repulsive screened electrostatic repulsions (19). Addition of non-adsorbing polymer induces attractive depletion interactions that lead to assembly of colloidal membranes, equilibrium structures consisting of a one-rod-length-thick monolayer of aligned rods with a fluid-like internal structure (12). For our experiments we use wild-type filamentous virus (*fd-wt)* and *fd*-Y21M that differs from its wild-type counterpart by a point mutation in the major coat protein (20, 21). Both viruses have comparable contour length (22), however, studies of bulk cholesteric phase demonstrates that *fd-wt* forms a left-handed cholesteric structure, whereas *fd*-Y21M forms a right-handed one (Fig. 1a). Bulk liquid crystalline studies also demonstrated that *fd-wt*/*fd*-Y12M mixture forms a homogeneous cholesteric phase with a pitch that depends on the virus ratio, $x_{fd}$ (21). The associated twist wave-number varies monotonically and smoothly from positive (right-handed) to negative (left-handed), changing sign at $x_{fd}$=0.26, the ratio at which the virus mixture is effectively achiral.

The structure of the colloidal membrane's edge is determined by the balance of the surface energy associated with the rod-depletion polymer interface and the elastic distortion energy originating from the non-uniform packing of rods within the membrane. The surface energy favors a curved edge profile, while elastic distortions favor a squared edge (15, 16, 18, 23). For *fd*-virus based colloidal membranes the surface energy dominates; consequently, the membrane's edge is curved and the edge-bound rods have to twist away from the membrane normal in order to fit the rounded profile imposed by the surface tension. Furthermore, the chirality of the filaments does not change the structure of the edge profile such as the twist penetration length, $\lambda_t$, just the effective edge tensions proportional to the energy difference between particles located at the edge and in the bulk (18).

The tilting of edge-bound rods away from the membrane normal results in structural and optical anisotropy in the *x*-*y* plane (Fig. 1b-c) (15, 16, 18). The optical anisotropy can be quantified by 2D-LC-PolScope that yields images where each pixel's intensity is proportional to the 2D projection in the *x*-*y* plane of the retardance, *R* (24). The resulting twist at the edge penetrates into the membrane interior over a characteristic length scale (17). A radial retardance profile yields a twist penetration length that is significantly different between *fd*-Y21M and *fd*-*wt* (Fig. 1f). However, the 2D projection of the retardance map does not reveal the handedness of the edge twist. To extract this information we use 3D-LC-PolScope (25). Briefly, a micro-lens array is introduced into the back-focal plane of the objective of the 2D-LC-PolScope, producing a grid of conoscopic images on the CCD camera. Each conoscopic image determines the local orientation of rods. An azimuthally symmetric retardance profile with a dark spot in the center indicates that rods at that locality are oriented along the z-axis. A shift of the zero-retardance spot away from the center of a conoscopic image yields the magnitude of the local virus tilting, while its radial position indicates the 3D direction of the birefringence vector. 3D-LC-PolScope images show that *fd-wt* membranes comprised of *fd-wt* and *fd*-Y21M viruses are right- and left-handed, respectively (Fig.1d-e).

## 2. Weakly chiral rod mixtures lead to scalloped membranes

Next, we examine the structure of colloidal membranes assembled from a mixture of *fd-wt* and *fd*-Y21M. The difference in their contour length of less than a few percent is not sufficient to induce lateral phase separation (26); instead we observe uniformly mixed membranes throughout the entire range of parameters studied here (Fig. 2). The binary mixture is characterized by the areal fraction $x_{fd} \equiv n_{fd}/(n_{fd}+n_{fdY21M})$ where $n_{fd}$ and $n_{fdY21M}$ are the concentration of *fd-wt* and *fd*-Y21M rods, respectively. The membranes are stable for a wide range of depletant concentrations and for all ratios $x_{fd}$ (Fig. 2a). The twist at the membrane edge is right-handed at low $x_{fd}$ and left-handed at high $x_{fd}$ . Surprisingly, for intermediate $x_{fd}$, ($0.04<x_{fd}<0.45$), we no longer observe flat circular membranes, instead the membrane's entire edge becomes decorated with a series of outward protrusions that are terminated by cusp-like defects (Fig. 2b). Furthermore, *z*-scans indicate that such scalloped membranes are not flat but have a distinct 3D structure where a cusp point defect located below the membrane is always followed by a defect located above the same plane (Fig. 3).

2D-LC-Polscope images of the scalloped membranes demonstrate that rods at the edge of each outward protrusion have the same twist penetration length (Fig. 3a-b, Fig. S1a). However, 3D-LC-Polscope reveals that adjacent protrusions have alternating left- and right-handed twist (Fig. 3c). The regions of opposite twist are separated by cusp-like point defects. Because each protrusion is always accompanied with two adjacent point defects alternating above and below the monolayer plane, there can only be an even number of cusp defects along the membrane circumference. The combined *z*-stack and 3D-LC-PolScope images allow us to schematize the edge structure of the scalloped membranes, which is more intricate when compared to the edge structure of chiral colloidal monolayers studied previously (Fig. 3d-f). The formation of the scalloped membranes is the direct consequence of molecular chirality, since scalloped membranes appear in the limit of weak chirality, i.e. between $0.04<x_{fd}<0.4$ (Fig. 2a).

In principle, there could be localized demixing of the two rods species, where *fd-wt* would preferentially localize at the edges with a left-handed twist, and *fd*-Y21M at the edges with opposite twist. This was not observed experimentally. We labeled all *fd-wt* rods with a fluorescent dye (Alexa 488) and *fd*-Y21M rods with fluorescent dye (Dylight 550). Using dual-view fluorescence, a technique that allows us to simultaneously image *fd-wt* and *fd*-Y21M fluorescent rods, we observe that, within experimental error, membranes in bulk and at the edges remain homogeneously mixed for all $x_{fd}$, even within the outwards protrusion (Fig. 2b, Movie 1-2). In a second series of experiments, we measured the twist penetration length, $\lambda_t$, the interfacial tension $\gamma$, and the edge bending rigidity $k_b$. For scalloped membranes, we find that outward protrusions with either handedness had the same $\lambda_t$ (Fig. S2) and $\gamma$ and $k_b$ that could not be distinguished within experimental error. Additionally, these quantities varied continuously from $x_{fd}$=0 to $x_{fd}$=1, which also indicates mixture homogeneity (Fig. S1).

## 3. Membrane coalescence generates cusp-like deformations.

Lateral coalescence of colloidal membranes can lead to the formation of unconventional defect structures. For example, two laterally coalescing membranes of the same handedness can trap 180 degrees of twist, resulting in a π-wall line defect (23). In order to elucidate a possible mechanism that leads to the formation of cusps in scalloped membranes, we observed membrane coarsening by using an angled-light 2D-LC-PolScope. This technique differs from conventional 2D-LC-PolScope; instead of having the light source aligned with the *z*-axis, the almost closed aperture associated with the back focal plane is translated away from the optical center, resulting in the plane waves illuminating the sample at an angle. This in turn reveals the handedness of the local rod twisting. We define a coordinate system in which the optical axis lies along the $z$-direction, and the membrane lies in the $x$-$y$ plane (Fig. 4a). The aperture of the condenser back focal plane is placed so that the incident illumination is tilted in the $x$-$z$ plane. It follows that the rods within a membrane along the $y$-axis (dashed line in Fig. 4b) exhibit a variable tilt with the respect to the illumination plane. The regions where rods are perpendicular to the plane of the illuminating wave will exhibit no optical retardance, while retardance will increase with an increasing tilt of rods away from the angle of the incident light. As a result, the lower edge of a right-handed membrane exhibits reduced retardance (relative to the rods in the bulk) as the viruses tilt towards the light source, whereas the upper edge exhibits increased retardance due to the rods tilting away from the light source (Fig. 4b). By the same reasoning, a left-handed membrane will exhibit the opposite behavior, with darker and brighter regions at the top and bottom of the membrane along the y-axis, respectively (Fig. 4c). This technique allows us to distinguish between left and right-handed membrane edges with a higher spatial resolution than 3D-LC-PolScope.

For $0.04<x_{fd}<0.45$, in the early stages of the sample maturation we observe circular membranes of either edge handedness, indicating spontaneously broken achiral symmetry. Over time, the intermediate sized membranes with mixed edge twist continue to coalesce. When two membranes with the same handedness merge, we observe the formation of either a $\pi$-wall or an array of pores at the coalescence junction, as was discussed previously (Fig. 4b, Movie 4) (23). By contrast, as the two proximal edges of a membrane pair with the opposite twist rupture, the adjoining neck widens and the twist of the edge-bound rods is expelled by aligning constituent rods with the membrane normal (Fig. 4c, Movie 3). This coalescence process leads to a daughter membrane that has two outward protrusions and two cusp defects at which the twist of edge-bound rods switches handedness. Once formed, the cusp defects remain stable indefinitely. An outward protrusion with a pair of defects can also be imprinted into the edge using optical tweezers (Fig. 4d, Movie 5). The method, which allows for robust engineering of cusp defects, consists of pulling the twisted ribbons out of the membrane edge and subsequently dragging it into the membrane.

## 4. Effective interactions between adjacent point defects

The structure of the scalloped edges greatly depends on the areal fraction $x_{fd}$ and how close the membrane is to the achiral limit ($x_{fd}$=0.26). At the boundary of stability of scalloped membranes, near $x_{fd}$=0.04 or 0.45, a pair of point defects remains bound to each other at a well-defined distance (Fig. 5a). In comparison, close to the achiral limit the defect pair freely moves along the edge and the total circumference of each outward protrusion exhibits significant fluctuations (Movie 6). These observations can be explained by the chiral control of membrane line tension (18). Increasing the rod chirality raises the free energy of the untwisted interior rods, while lowering the free energy of edge bound twisted rods, thus leading to the chiral control of the line tension. Likewise, chirality can also raise the line tension if the twist at the membrane's edge is the opposite of the natural twist preferred by the constituent molecules.

For achiral membranes the line tension associated with the exposed edge of left-handed and right-handed outward protrusions is roughly equal. The overall free energy does not significantly change as one outward protrusion extends its length at the expense of another one, by translating the cusp defect. In this limit the point defects freely diffuse and the lengths of outward protrusions with either handedness exhibit significant fluctuations in agreement with experimental observations. However, away from the achiral limit there is a finite difference in line tension between the left-handed and right-handed outward protrusions, and the free energy is minimized by reducing the length of the outward protrusions with unfavorable twist.

In order to quantitatively test these ideas we have measured the effective interaction between a pair of point defects that are connected by a single protrusion. We used phase contrast microscopy to track the positions, $s_i$, of two adjoining defects along the membrane contour (Fig. 5a). For an achiral sample ($x_{fd}$ = 0.26), the separation between two adjoining defects, $\delta s=s_{i+1}-s_i$, fluctuates by many microns over a timescale of minutes (Fig. 5b). However, away from achiral limit, we observe that the relative separation between these defect pairs remains well defined on experimental time scales. We measured the probability distribution function, $P(\delta s)$, of the defects being separated by distance $\delta s$. The measured distributions are described by a Gaussian: $P(\delta s)=\exp(-\alpha(\delta s- \delta s_0)^2/2k_BT)$, indicating that the defects are bound by a harmonic potential centered around the equilibrium separation, $\delta s_0$ (Fig. 5c). The equilibrium defect separation as well as the strength of the effective binding potential, α, depends on $x_{fd}$, the ratio of left- and right-handed rods. By varying membrane composition we extracted how α, as well as the equilibrium separation, $\delta s_0=<\delta s>$ depends on $x_{fd}$ (Fig. 5d, 5e). Approaching the achiral mixture limit ($x_{fd}$ = 0.26) leads to the divergence of the mean separation $\delta s_0$ and a vanishing α. In this limit the adjacent defects effectively decouple from each other. Increasing chirality away from the achiral limit decreases equilibrium separation and increases the coupling strength, indicating tighter defect binding. The existence of a finite equilibrium separation indicates a competition between short-range repulsion, due to elastic distortions, and long-range attraction caused by the asymmetry of the line tension associated with the edges of the opposite twist**.**

The passive fluctuation analysis only maps the binding potential within a few $k_BT$ around its minimum. To measure the entire binding potential we performed active experiments where we moved one defect by $\delta s$ using an optical trap, while simultaneously measuring the force $F$ exerted on the other defect (Fig. 6a, 6b). For this purpose, we embedded 1.5 μm diameter colloidal beads into two adjoining cusp defects. Once placed there, beads remained attached to a defect for the entire duration of the experiment. To ensure that the beads do not alter the defect structure we measured thermal fluctuations of a defect pair with and without embedded beads and found them to be identical within experimental error (Fig. S3). We then calibrated the trap to measure the zero force at the equilibrium distance $\delta s_0$ (Fig. S4) and determined the optimal laser power to measure the force $F$ (Fig. S5). We extracted the force as a function of $\delta s$, $F(\delta s)$, which is averaged over 10 identical experiments (Fig. 6a, 6b). In the vicinity of the equilibrium separation, $\delta s_0$, the force measurements quantitatively agree with the fluctuation experiments described above. As expected, the force is negative below $\delta s_0$, and positive above $\delta s_0$, confirming that $\delta s_0$ is the stable equilibrium position. The force steeply increases for small separations and saturates at large separations, indicating that a defect pair is permanently bound. The magnitude of the force plateau and the slope of the force-increasing region depend on the areal fraction $x_{fd}$. By moving farther away from the achiral limit, we find that the equilibrium distance between bound defects $\delta s_0$ decreases. These experiments also demonstrate that the pairwise defect interactions are governed by a balance between short-range repulsion and long-range attraction.

## 5. Modeling the interactions between two adjacent point defects

To investigate the stability of a scalloped membranes and the interactions between adjacent point defects, we have developed an elastic theory based on the Helfrich free energy of fluid membranes. Since there is an axis of reflection at the midline between adjoined cusps, our model reduces the overall 3D geometry of the membrane to an isolated configuration around a point defect (27). The outward protrusions between two neighboring cusps must form via the interplay between the line tension $\gamma$, the interfacial bending rigidity $k_b$, and the variables associated with the overall membrane deformation. For an isolated defect, the free energy is then given by (27):

$$F_2 = \int dS\left(\bar{k}\kappa_G + \sigma\right) + \sum_{i=1,2} \oint ds_i\left(\gamma_i + k_b\kappa_{s,i}^2\right), \qquad \text{(Eq. 1)}$$

where $\kappa_G$ is the Gaussian curvature, $\bar{k}$ is the corresponding elastic modulus (28), and $\sigma$ denotes the surface tension. For simplicity, we will assume that the mean curvature, $H$, vanishes in scalloped membranes. Consequently, $\sigma$ does not contribute to the force balance on the surface. The bulk terms are integrated over the membrane area with an element $dS$, whereas the interfacial terms are integrated along the arc length with elements $ds_i$. The two edge profiles (enumerated by the index $i$) with opposite handedness meet at the cusp defect and are in general different, since their curvature $\kappa_{s,i}$ and line tension $\gamma_i$ can be unequal. The relaxation length of each edge from a space curve to a straight line in the monolayer plane is given by the natural length scales $\xi_i \equiv \sqrt{k_b / \gamma_i}$ ($\xi_i \sim 0.5$ μm). The minimization procedure of Eq. (1) by a variational analysis is discussed elsewhere (27).

The structure of the scalloped edge is determined by the balance between two contributions to the free energy, i.e. the line energy and the surface energy. On the one hand, the line energy suppresses the formation of outward protrusions and cusp defects, since they increase the total membrane circumference. On the other hand, each cusp defect generates negative Gaussian curvature, which lowers the free energy of elastic deformations if the Gaussian modulus is positive and sufficiently large (27). Based on the interplay between these two contributions, our model predictsregions where scalloped membranes are more stable than flat circular membranes as a function of $\bar{k}$ and $x_{fd}$ (Fig. 6c). To calculate this phase diagram we have used experimental values for $\gamma_1$ and $k_b$ (Fig. S1). When the areal fraction of the virus mixture deviates from $x_{fd}$ =0.26, an increasing magnitude of $\bar{k}$ is required to stabilize scalloped membranes (Fig. 6c). The reason is that away from the achiral limit, one of the edges (e.g. associated with $\gamma_2$) has incompatible chirality with the preferred overall handedness along the membrane boundary. Consequently, the rods along that edge tilt into a high-energy configuration, as opposed to the molecules in the adjacent outward protrusion that has lower energy (with $\gamma_1$). This leads to $\Delta\gamma \equiv \gamma_2 - \gamma_1 > 0$, i.e. the overall free energy of the scalloped membranes rises, and a long-range attractive interaction between two adjoined defects emerges. The two defects, however, cannot approach very close to each other since the surface in between must then flatten, leading to a diminishing negative Gaussian curvature that in turn raises the free energy. This yields short-ranged repulsive interactions between nearby defects (27, 29). The equilibrium defect separation is determined by the competition between these two effects.

Theoretical predictions for the equilibrium separation of defects, $\delta s_0$, and their coupling strength, $\alpha$ were fit to experimental measurements (Fig. 5d, 5e). The line tension $\gamma_1$ and the bending rigidity $k_b$ where taken from experiments (Fig. S1), while we took $\Delta\gamma$ and $\bar{k}$ as fitting parameters. The power spectrum of the membrane edge fluctuations yields $\gamma_1$ and $k_b$ within ~ 10% error. However, since $\gamma_2$ is very close to $\gamma_1$, measurements are not precise enough to extract the effective difference between them, so $\Delta\gamma$ remains a free parameter. The fitting procedure yields $\Delta\gamma$ as a polynomial $\Delta\gamma = (-115 x_{fd} + 30) k_B T / \mu m$ that vanishes at $x_{fd}$ =0.26 and becomes 30 $k_B T/\mu m$ at $x_{fd}$ =0, i.e. $\Delta\gamma$

stays within the bounds of experimental uncertainty. Likewise, theoretical fits to experimental curves yield the magnitude of the Gaussian modulus, $\bar{k} = 200 k_B T$ .

The theory quantitatively reproduces how the effective defect interactions (coupling strength α and equilibrium separation $\delta s_0$) depend on $x_{fd}$, thus confirming their origin: a long-range attraction due to $\Delta\gamma>0$, and a short-range repulsion associated with membrane Gaussian curvature (Fig 5d, 5e). The modulus α as a function of $x_{fd}$, extracted from a linear approximation to the theoretical force-extension curves at the point where the force vanishes, qualitatively agrees with the experimental profiles (Fig. 5d). We note that, without the Gaussian curvature contribution, a simpler 2D theory modeling a flat and thermodynamically unstable scalloped membrane consistently yields smaller $\delta s_0$ values than those from the 3D model presented here (Fig. S6). Hence, the Gaussian curvature term with a positive modulus explains the nature of in-plane and out-of-plane deformations as well as the overall stability of the scalloped membranes. Furthermore, the model with the same parameters also reproduces the optical tweezer measurements of the effective defect interactions over a much larger range of separations (Fig. 6b).

Certain precautions need to be taken when interpreting the extracted magnitude of the Gaussian modulus $\bar{k} = 200 k_B T$ . In particular, another prediction of the model is that $\bar{k}$ is equal to the product of the in-plane protrusion amplitude $A$ (see Fig. 3f) and the line tension (27). This is because bigger protrusions would make the edge longer at constant γ, necessitating a higher $\bar{k}$ to stabilize the scalloped membranes. At $x_{fd}$ =0.26, when the protrusion size is 2 μm (see Fig. 3a) and the line tension is $\gamma \sim 500 k_B T / \mu m$ (see Fig. S1d), the Gaussian modulus from this relation is found as $\bar{k} = \gamma A \sim 10^3 k_B T$ . This value is almost an order of magnitude higher than $\bar{k} = 200 k_B T$ extracted for the theoretical fits (Fig. 5d, 5e). The resulting discrepancy between two estimates of the Gaussian modulus may be due to the fact that our model relies on a simple geometrical assumption, an axially symmetric catenoidal surface, which likely accumulates more negative Gaussian curvature than the experimental shape of the membrane surface. Therefore, our analysis underestimates the Gaussian modulus that stabilizes the scalloped membrane over a flat configuration. Theoretically, compromising axial symmetry or the smoothness of the surface around the cusp could yield a minimal saddle surface with a lower amount of the total Gaussian curvature. On the one hand, this would restore $\bar{k}$ to higher values to stabilize the scalloped membranes and resolve the discrepancy between two estimates. On the other hand, the complexity of our model would then greatly increase. We note that, experimentally, the surface around the cusp must be governed by the local matching of the rod orientations, which may indeed form a non-smooth surface at the cusp. In this configuration, the structural relation between the Gaussian modulus, the protrusion amplitude, and the line tension must still hold, as discussed elsewhere (27). Nevertheless, our theory in its current simplified form still gives an accurate description of the scalloped membrane structure and stability.

## 6. Theoretical Estimate of Gaussian Curvature Modulus

We can use a simple argument to estimate the Gaussian curvature modulus, $\bar{k}$, of colloidal membranes of thickness *D*, surrounded by a depleting polymer with effective diameter $d$. There are two distinct contributions to $\bar{k}$: an intrinsic contribution arising from the internal stresses among the virus particles $\bar{k}_m$, and an entropic contribution arising from the polymer depletants $\bar{k}_p$. First, we consider the intrinsic contribution. Because it is fluid, we assume that the membrane middle surface membrane does not stretch when the membrane bends. The rods can adjust around each other to accommodate a change in curvature without changing their equilibrium spacing in the midplane. However, imposing any curvature onto the membrane will induce a strain that depends on *z*, the distance along the membrane normal away from the midplane. In particular if the membrane has mean curvature H and Gaussian curvature $\kappa_G$, then the areal strain of the surface at distance $z$ is given by $\varepsilon(z) = 2Hz + \kappa_G\, z^2$ (30). The corresponding lateral membrane stress is $\sigma(z) = \sigma_0(z) + Y\varepsilon(z)$, where $\sigma_0(z)$ is the stress in the membrane when it is flat, and $Y$ is the modulus for areal compression. Since the rods are uniform along their lengths we take $Y$ to be independent of *z*. The lateral stress is isotropic because the membrane is fluid. There is a compressive stress in the membrane even when it is flat because the polymer depletants squeeze the membrane (31). The total volume excluded to the polymers for a flat membrane of area $A$ is $V_{0,\mathrm{ex}} = A(D + d)$, leading to a contribution to the free energy per unit area $\gamma = nk_BT(D+d)$, which can also be considered an entropic tension (16). To balance this tension, the rods must experience a compressive stress $\sigma_0(z) = -n\, k_BT(D+d)/D$. To calculate the contribution to $\bar{k}$ from the rods in the membrane with zero mean curvature, we write the membrane free energy per unit mid surface by integrating the stress with the respect to the strain as $\mathcal{F}_m = \int_{-D/2}^{D/2} dz \int_0^{\varepsilon(z)} d\varepsilon' \sigma(\varepsilon') = \bar{k}_m \kappa_G$, where $\bar{k}_m = \int_{-D/2}^{D/2} dz\, z^2 \sigma_0(z)$ (32, 33). The intrinsic contribution to the Gaussian curvature modulus is negative, $\bar{k}_m = -nk_BT(D+d)D^2/12$. Now consider the polymer contribution to the free energy, arising from the excluded volume of a curved membrane with zero mean curvature, $V_{ex} = \int dA \int_{-(D+d)/2}^{(D+d)/2} (1+\kappa_G\, z^2) dz$. This equation demonstrates that imposing Gaussian curvature onto a membrane reduces volume that is excluded to the depleting polymers and thus increases the system free energy. This contribution is positive and more significant when compared to contribution due to internal membrane stresses. Integrating across the membrane thickness and using $F_p = nk_BTV_{ex} = \gamma A + \bar{k}_p \int dA\kappa_G$ yields $\bar{k}_p = (D+d)^3 nk_BT/12$. yields $\bar{k}_p = (D+d)^3 nk_BT/12$. The net Gaussian curvature modulus is therefore $\bar{k} = \bar{k}_p + \bar{k}_m \approx D^2 dnk_BT/6$, where the approximation follows since the membrane is much thicker than the polymer particles, $D >> d$.

To estimate the polymer contribution to the Gaussian curvature modulus, we use *n*~40 mg/mL, *D*~880 nm and *d*~30 nm for a Dextran with molecular weight 500,000 g/mol (34). With these numbers, we find that $\bar{k}$ ~ 185 $k_BT$. Despite its approximate nature, our estimate yields the magnitude $\bar{k}$ that is in reasonable agreement with experimental measurements. Note that our argument for the polymer contribution is similar to the model put forward to explain negative Gaussian curvature modulus for surfactant interfaces: for a saddle-splay surface with *H*=0, there is less room for the surfactant chains, which therefore must stretch and incur a higher free energy (35).

## 7. Discussions and Conclusions

Our combined theoretical and experimental work demonstrates that membranes comprised of achiral rods exhibit higher structural complexity when compared to flat membranes assembled from chiral rod-like viruses. In the latter case strong chirality enforces uniform twist of rods along the entire membrane circumference, leading to the formation of flat 2D disks. By contrast, weakly chiral or achiral membranes exhibit an intriguing instability that is driven by an interplay between the Gaussian curvature of a colloidal membrane and the spontaneous achiral symmetry breaking of rods located at the membrane's edge. The achiral symmetry breaking induces formation of cusp-like defects. These defects in turn allow the membrane to adopt a three dimensional shape that decreases the overall energy associated with its negative Gaussian curvature.

Despite the important role it plays in diverse processes, measuring the Gaussian modulus of conventional lipid bilayers remains a significant experimental challenge. In comparison, the properties of the colloidal membranes described here allow us to estimate their Gaussian modulus. Conventional bilayers have a negative Gaussian modulus, which means that saddle-shaped deformations increases the membrane energy (10, 11, 36, 37). On the contrary, experiments described here, as well as previous observations of diverse assemblages with excess Gaussian curvatures such as arrays of pores and twisted ribbons (18, 23), demonstrate that colloidal monolayers, in contrast to lipid bilayers, have positive Gaussian moduli.

Achiral symmetry breaking has been observed in diverse soft systems with orientational order, ranging from lipid monolayers and nematic tactoids to confined chromonic liquid crystals (38-44). In particular the measured structure and interactions of the cusp-like defects in colloidal membranes resemble studies of point defects moving along a liquid crystalline dislocation line in the presence of chiral additives (45). The main difference is that in the colloidal membranes the achiral symmetry breaking leads to out-of-plane 3D membrane distortions that couples liquid crystal physics to membrane deformations. This is not possible for inherently confined liquid crystalline films.

From an entirely different perspective, a number of emerging techniques have been developed to fold, wrinkle, and shape thin elastic sheets with in-plane elasticity (46-49). So far these efforts were focused on studying instability of thin elastic films with finite in-plane shear modulus. The methods to achieve folding or wrinkling of thin sheets involves either engineering of in-plane heterogeneities or imposing an external force. Our work demonstrates that simpler uniform elastic sheets lacking in-plane rigidity can spontaneously assume complex 3D folding patterns that decorate its edge.

Finally, methods described here and in our previous work should be applicable to any monodisperse rod type with sufficiently large aspect ratio. Thus, they might offer a scalable method for robust assembly of photovoltaic devices comprised of nanorods. Our previous investigation of chiral *fd-wt* colloidal membranes demonstrated that the twist at their edges introduces a significant energetic barrier that suppresses their lateral coalescence (23). In such samples membranes with diameters ranging from 10 to 100 μm are commonly found. Compared to chiral colloidal membranes, we find that colloidal membranes of monodisperse virus mixtures that are close to the achiral limit coalesce much faster and can easily reach millimeter dimensions.

**Acknowledgements:** T.G. acknowledges the Agence National de la Recherche Française (ANR-11-PDOC-027) for support. P.S., C.N.K, R.B.M. and Z.D acknowledge support of National Science Foundation through grants: MRSEC-1420382, NSF-DMR-0955776 and NSF-DMR-1609742. We also acknowledge use of Brandeis MRSEC optical and biomaterial synthesis facility supported by NSF-MRSEC-1420382. T.R.P. acknowledges support of the National Science Foundation through MRSEC-1420382 and NSF-CMMI-1634552. We also acknowledge conversations with William Irvine, Robert Pelcovits and Leroy Jia. R.D.K. was partially supported by a Simons Investigator grant and NSF-DMR-1262047.

**Author contributions:** T.G. and Z.D. designed the project. T.G., P.S., A.W. and M.J.Z. performed the experiments. C.N.K. and R.B.M. developed the theoretical model explaining the structure and interactions of the scalloped membranes. T.G. and C.N.K. analyzed the results. T.G., C.N.K., M.J.Z., and Z.D. wrote the manuscript. R.O. performed 3D LC-PolScope image acquisition. R. K. and T. P. explained the Gaussian modulus of colloidal membranes. All authors reviewed the manuscript.

## Supplementary material

**Sample Preparation:** Both viruses, *fd* and *fd*-Y21M were grown in bacteria and purified as described elsewhere (18). *fd*-Y21M, has a single point mutation in the amino-acid sequence of the major coat protein: amino acid number 21 is replaced from Y to M. *fd* and *fd*-Y21M were labeled with fluorescent dye as described elsewhere (50). The preparation of optical chambers was described elsewhere (18)

**Optical Microscopy:** Experiments were carried out on an inverted microscope (NikonTE 2000) equipped with traditional polarization optics, a Differential Interference Contrast (DIC) module, a fluorescence imaging module and 2D-LC-Polscope module. For dual view fluorescence imaging we used DV2 from Photometrics. We used a 100X oil immersion objective (PlanFluor NA 1.3 for DIC and PlanApo NA 1.4 for phase contrast). Images were recorded with cooled CCD cameras (CoolSnap HQ (Photometrics) or Retiga Exi (QImaging, Surrey BC, Canada)). For 3D-LC-PolScope measurements, we used a Zeiss Axiovert 200M microscope with a Plan Apochromat oil immersion objective (63X/1.4NA) and a monochrome CCD camera (Retiga 4000R, QImaging, Surrey BC, Canada).

**Laser tweezers:** A 1064 nm laser (Coherent Compass) was brought into the optical path of an inverted microscope (Nikon Eclipse Te2000-u) and focused with a 100X objective onto the image plane (Nikon PlanFlour, NA 1.3). To simultaneously trap multiple beads, a single beam was time shared between different positions using an acousto-optic deflector (IntraAction-276HD) (51). Bead position was measured using back focal plane interferometry and a quadrant photodiode (QPD) (52). A separate 830 nm laser (Point Source Iflex-2000) was used as a detection beam. To calibrate the photodiode we scan a bead across the detection beam in known step sizes and measure the corresponding voltage change. Trap stiffness was calibrated by analyzing the power spectrum of the bead position (52).

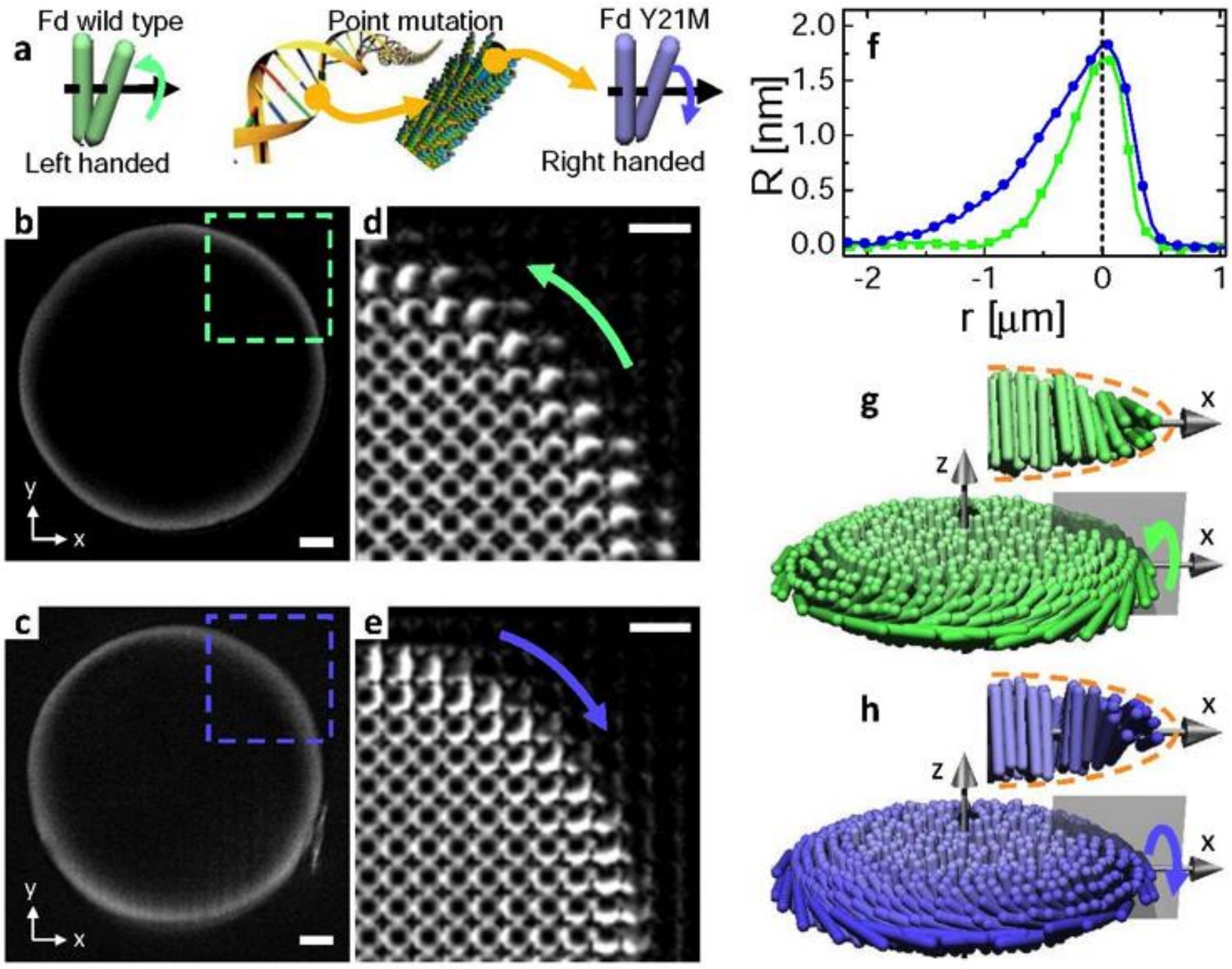

**Fig. 1 Microscopic chirality of constituent rods determines preferred twist at the membrane's edge. (a)** Bacteriophage *fd-wt* is a rod-like molecule with left-handed chirality (green). A single point mutation of the major coat protein switches the microscopic chirality yielding *fd*-Y21M (blue). **(b, c)** 2D-LC-PolScope images of *fd-wt* and *fd*-Y21M colloidal membranes. The local twist at the membrane's edge results in optical retardance that is visualized with polarization techniques. The retardance is coded in a linear grayscale that varies from $R$=0 nm (black) to $R$=3 nm (white). **(d, e)** 3D-LC-PolScope image of *fd-wt* and *fd*-Y21M membranes reveals the twist of the edge-bound rods is left-handed for *fd-wt* and right-handed for *fd*-Y21M. **(f)** Comparison of the radial retardance profile, *R(r),* for both *fd-wt* and *fd*-Y21M membranes. The membrane interior is located at $r$<0 and its edge is at $r$=0. For *fd-wt* membranes the twist penetration length is $\lambda_t$=0.45±0.05 μm, while for *fd*-Y21M

membranes it is $\lambda_t$=1.00±0.11μm. **(g, h)** Schematics of *fd-wt* and *fd*-Y21M colloidal membranes. Scale bars, 2 μm.

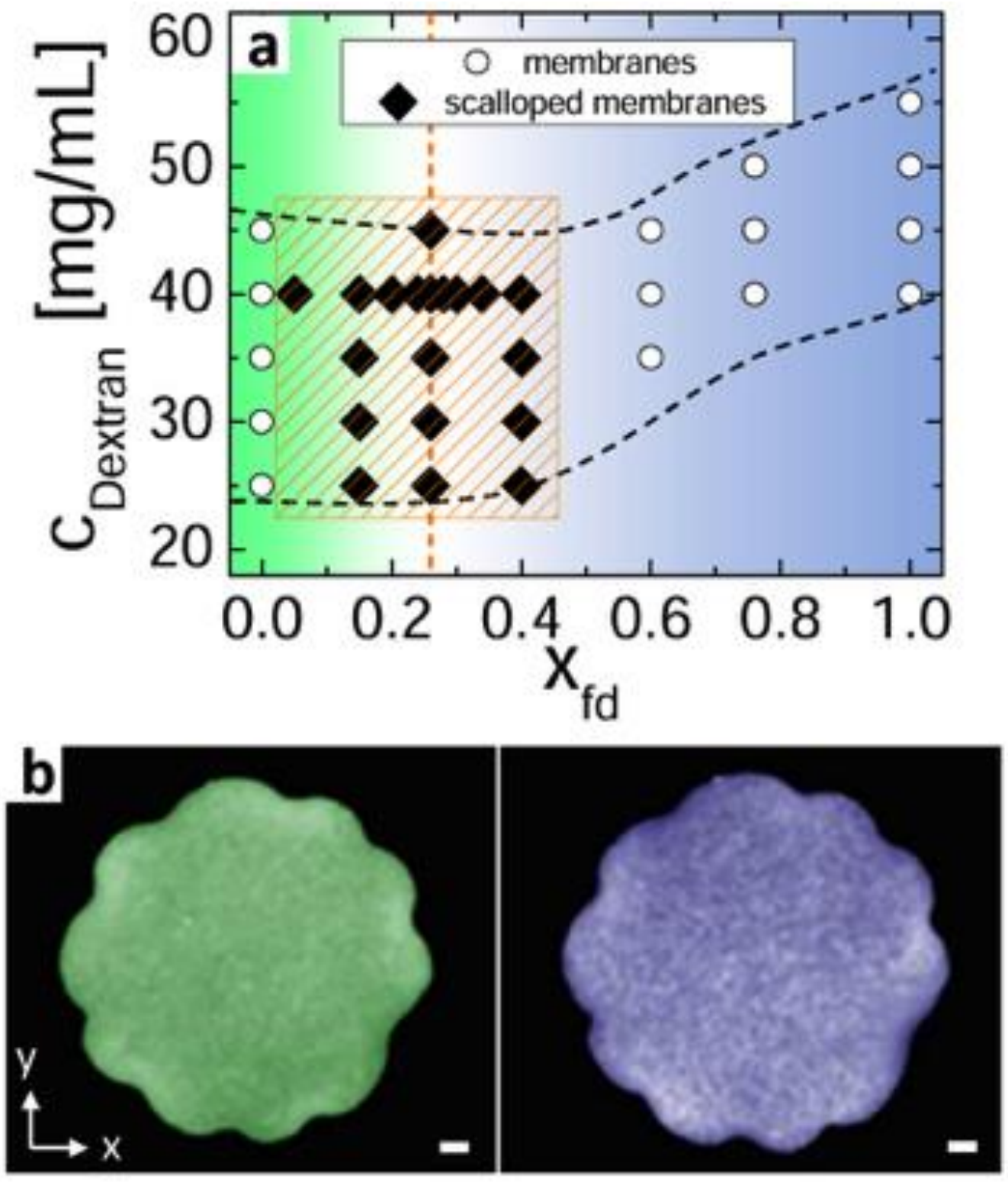

**Fig.2 Colloidal membranes assembled from *fd-wt*/*fd*-Y21M mixtures**. **(a)** Phase diagram of the *fd-wt*/*fd*-Y21M mixture as a function of the areal fraction $x_{fd}$ versus the depletant concentration $C_{Dextran}$. We observe scalloped membranes around the achiral ratio $x_{fd}$ =0.26 (vertical dashed line) in the orange shaded region. **(b)** Dual view fluorescence imaging of a scalloped membrane at $x_{fd}$ =0.26 and $c_{Dextran}$=40mg/mL. *fd-wt* viruses are labeled with the fluorescent dye Alexa-488 (left) and *fd*-Y21M viruses with DyLight-550 (right). The density distribution of both *fd-wt* (left) and *fd*-Y21M (right) is homogeneous throughout the membrane interior. Scale bars are 2 μm.

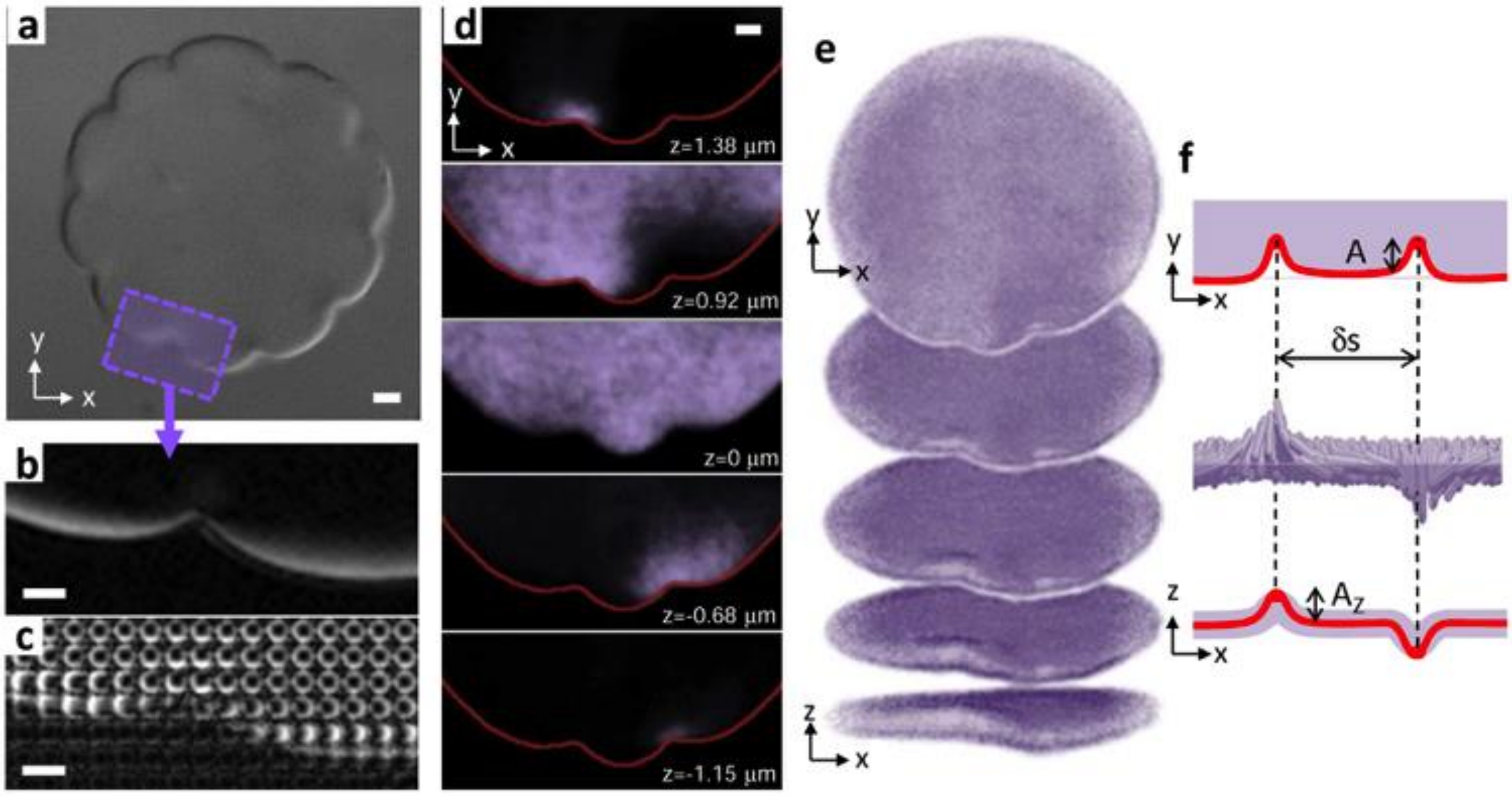

**Fig. 3 Structure of scalloped membranes**. **(a)** Differential Interference Contrast (DIC) image of a scalloped membrane formed in a *fd-wt* /*fd*-Y21M mixture at the achiral ratio $x_{fd}$=0.26 ($C_{dextran}$=40 mg/mL). The membrane's edge is decorated with a series of cusps separated by local outward protrusions. **(b)** 2D-LC-PolScope image of the membrane profile around a point defect. The twist penetration length $\lambda_t$ is identical on both sides of the cusp (Fig. S2). **(c)** 3D-LC-PolScope image of the membrane profile surrounding the defect. The viruses have opposite twist on either side of the point defect. **(d)** *Z*-scan of the scalloped membranes under confocal microscopy. The point defects alternate above and below the monolayer plane. **(e)** Reconstruction of the membrane based on the confocal images. **(f)** Schematic of a point defect inferred from the measurements in (c) and (d). The protrusion amplitude in the *xy*-plane is denoted by *A* and the cusp height by $A_z$. Scale bars are 2μm.

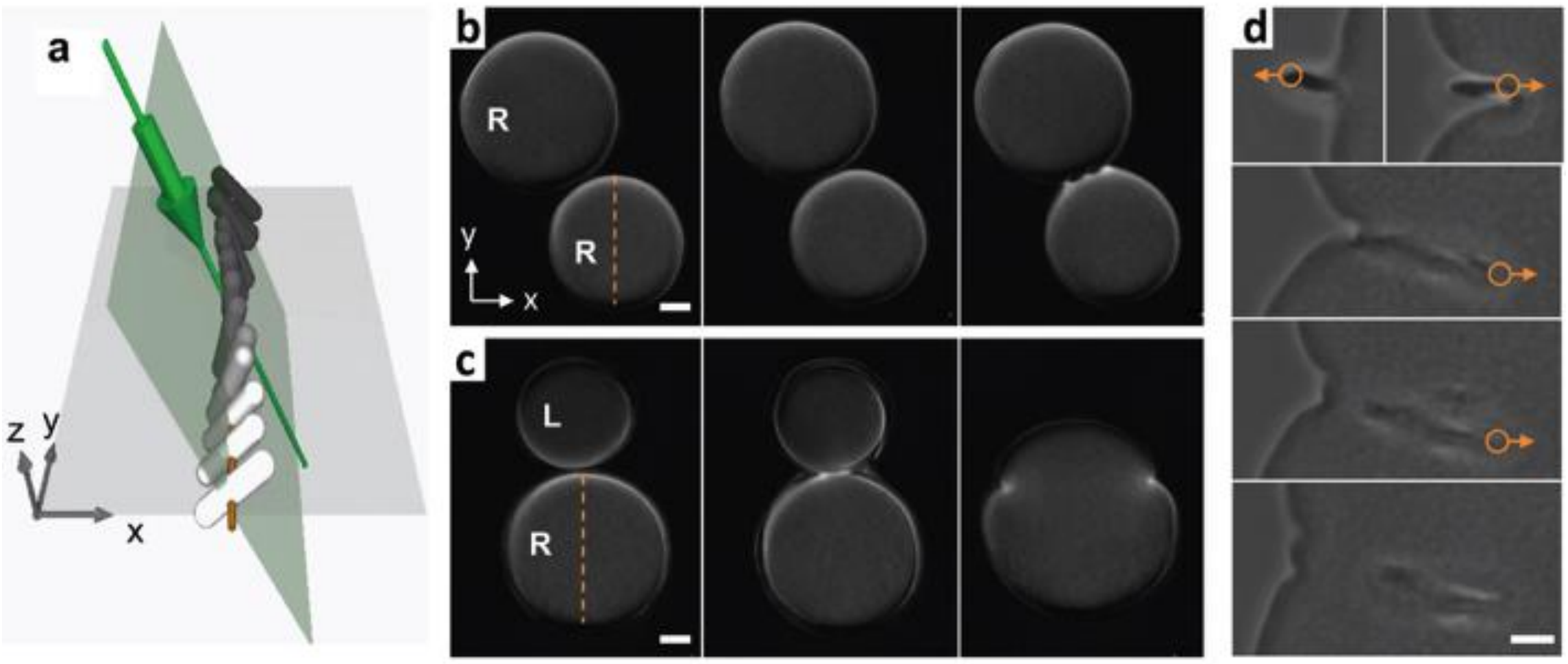

**Fig. 4. Lateral coalescence of membranes with opposite chirality leads to scalloped edges. (a)** Angled light illuminating 1D cross-section of twisted rods reveals the handedness of edge-bound rods. Because the direction of the incident light is tilted towards the *x* axis, rods twisting counter-clockwise along the light source are not birefringent and appear dark. In contrast, rods twisting clockwise, away from the incoming light, have higher optical anisotropy and thus appear bright. Tilting the light source breaks the symmetry of the 2D-LC-PolScope setup and allows us to distinguish between left-handed (L) and right-handed (R) membranes with a higher spatial resolution than that of the 3D LC PolScope. The grayscale changes from dark to light with increasing retardance, where the rods aligned with the direction of incident light have zero retardance. **(b)** The coalescence of two right-handed membranes, which are both bright at the top and dark at the bottom, results in the formation of pores. **(c)** Coalescence of a left-handed membrane with a right-handed membrane produces two cusps that separate the left-handed edge section from the right-handed one. **(d)** Artificial defects can be created with laser tweezers. Scale bars, 4 μm.

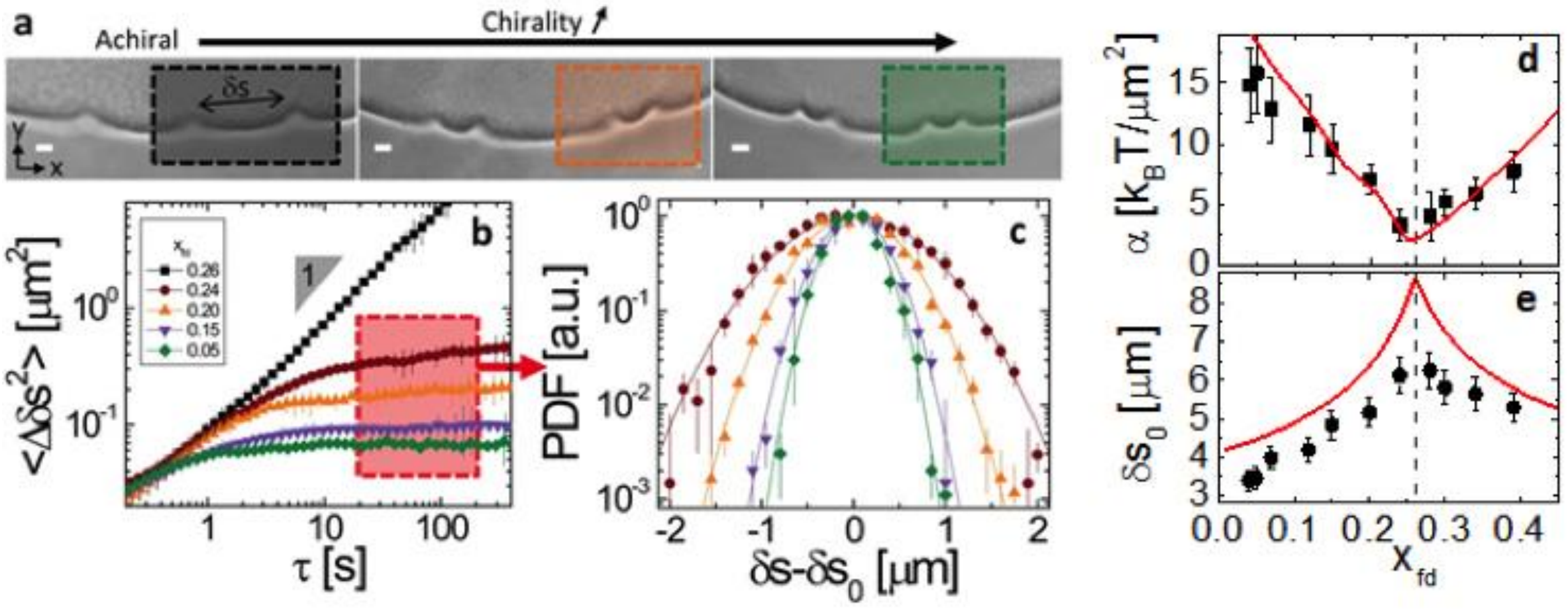

**Fig.5. Ratio of left- and right-handed rods tunes the effective defect interactions**. **(a)** Phase contrast image of scalloped membranes at $C_{Dextran}$ = 40 mg/mL. Areal fraction is decreasing from left to right ($x_{fd}$ =0.26, 0.15, and 0.05). Increasing the chirality by lowering the areal fraction $x_{fd}$ leads to tighter coupling between point defects due to the increasing difference in the line tension of the protrusions with left- and right-handed twist. Scale bars are 2 μm. The color of the shaded boxes indicates different MSD of defect separation in (b). **(b)** Mean square fluctuations of the separation of two coupled defects as a function of time at different $x_{fd}$. **(c)** Normalized probability distribution function, $P(\delta s\text{-}\delta s_0)$, extracted from relative separation of two defects for different values of $x_{fd}$ taken at $\tau \sim 100$ *s* where $<\Delta\delta s^2>$ plateaus (red box in (b)). $P(\delta s\text{-}\delta s_0)$ is fitted by a Gaussian distribution $\exp(-\alpha(\delta s\text{-}\delta s_0)^2/2k_BT)$. **(d)** The spring constant $\alpha$ of the harmonic potential around the equilibrium position $\delta s_0$ as a function of $x_{fd.}$ **(e)** Equilibrium distance between the defects $\delta s_0$ as a function of $x_{fd}$.

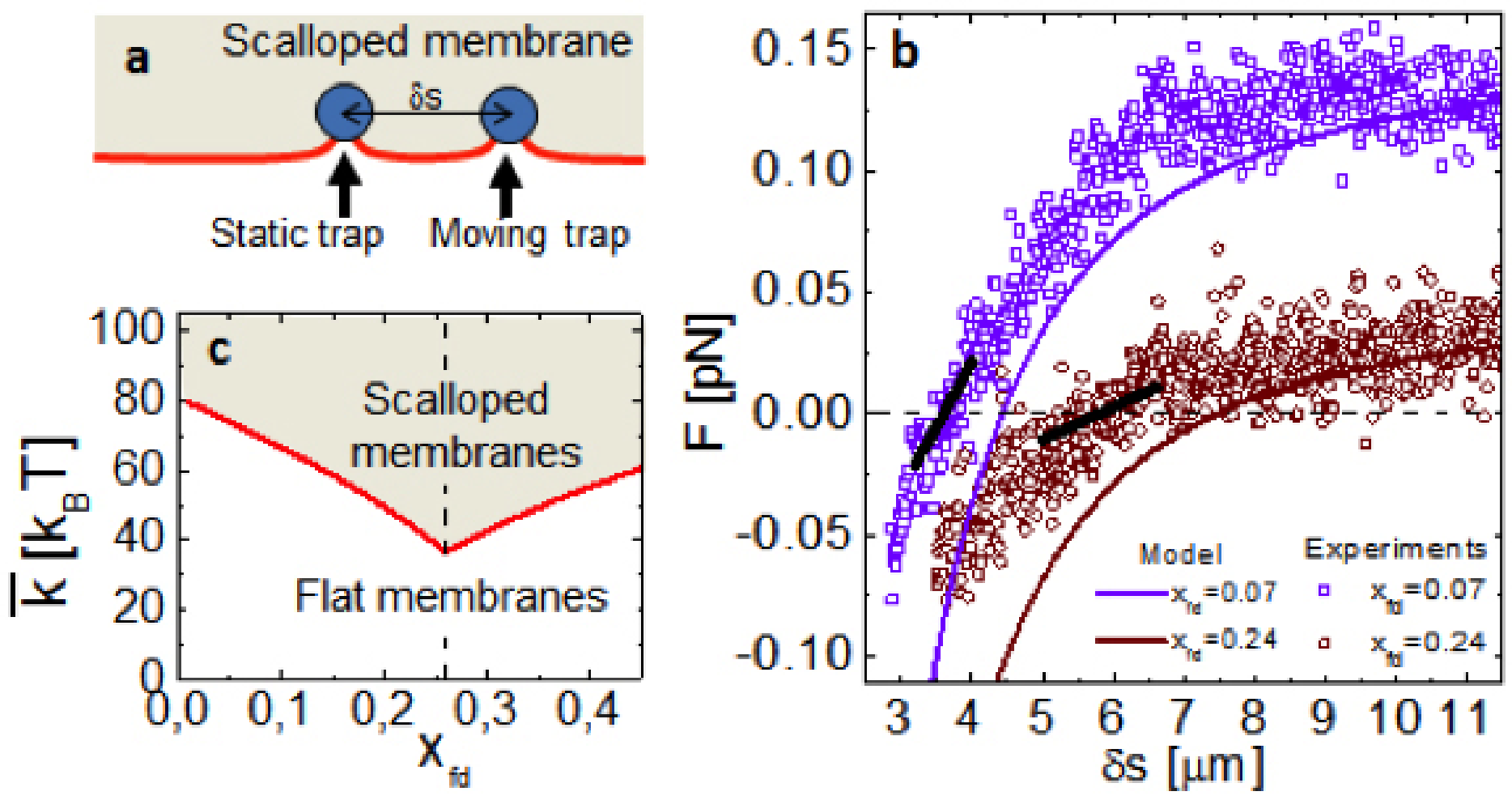

Fig. 6 **Measurement of defect binding and theoretical model of scalloped membranes.** (**a)** Schematic of the active experiments to measure defect interactions. Beads are embedded into defects. One defect is moved using an optical trap, and the force, *F*, exerted on the static defect is simultaneously measured. **(b)** Comparison between the force measurements (dots) obtained with laser tweezers and the theoretical model (full curves) as a function of δs, the distance between two adjacent defects. The black lines correspond to the measurements from the passive experiments (Fig. 5). **(c)** Theoretical phase diagram indicating the stability region of scalloped membranes as a function of the Gaussian modulus $\bar{k}$ versus areal fraction $x_{fd}$. The red line is the boundary between the flat and scalloped membrane phases.

# Achiral symmetry breaking and positive Gaussian modulus lead to scalloped colloidal membranes

Thomas Gibaud[1,2], C. Nadir Kaplan[1,3], Prerna Sharma[1,4], Andrew Ward[1,5], Mark J. Zakhary[1], Rudolf Oldenbourg[6], Robert B. Meyer[1], Randall D. Kamien[7], Thomas R. Powers[8,9] and Zvonimir Dogic[1]

[1]*The Martin Fisher School of Physics, Brandeis University, 415 South Street, Waltham, Massachusetts 02454, USA*
[2]*Univ Lyon, Ens de Lyon, Univ Claude Bernard, CNRS, Laboratoire de Physique, F-69342 Lyon, France*
[3]*John A. Paulson School of Engineering and Applied Sciences, Harvard University, Cambridge, MA 02138, USA*
[4]*Department of Physics, Indian Institute of Science, Bangalore 560012, India*
[5]*Program in Cellular and Molecular Medicine, Boston Children's Hospital, Boston, MA, USA*
[6]*Marine Biological Laboratory, 7 MBL Street, Woods Hole, MA 02543, USA*
[7]*Department of Physics and Astronomy, University of Pennsylvania, Philadelphia PA 19104, United States*
[8]*School of Engineering, Brown University, Providence, Rhode Island 02912, USA*
[9]*Department of Physics, Brown University, Providence, Rhode Island 02912, USA*

## Supplementary Figures

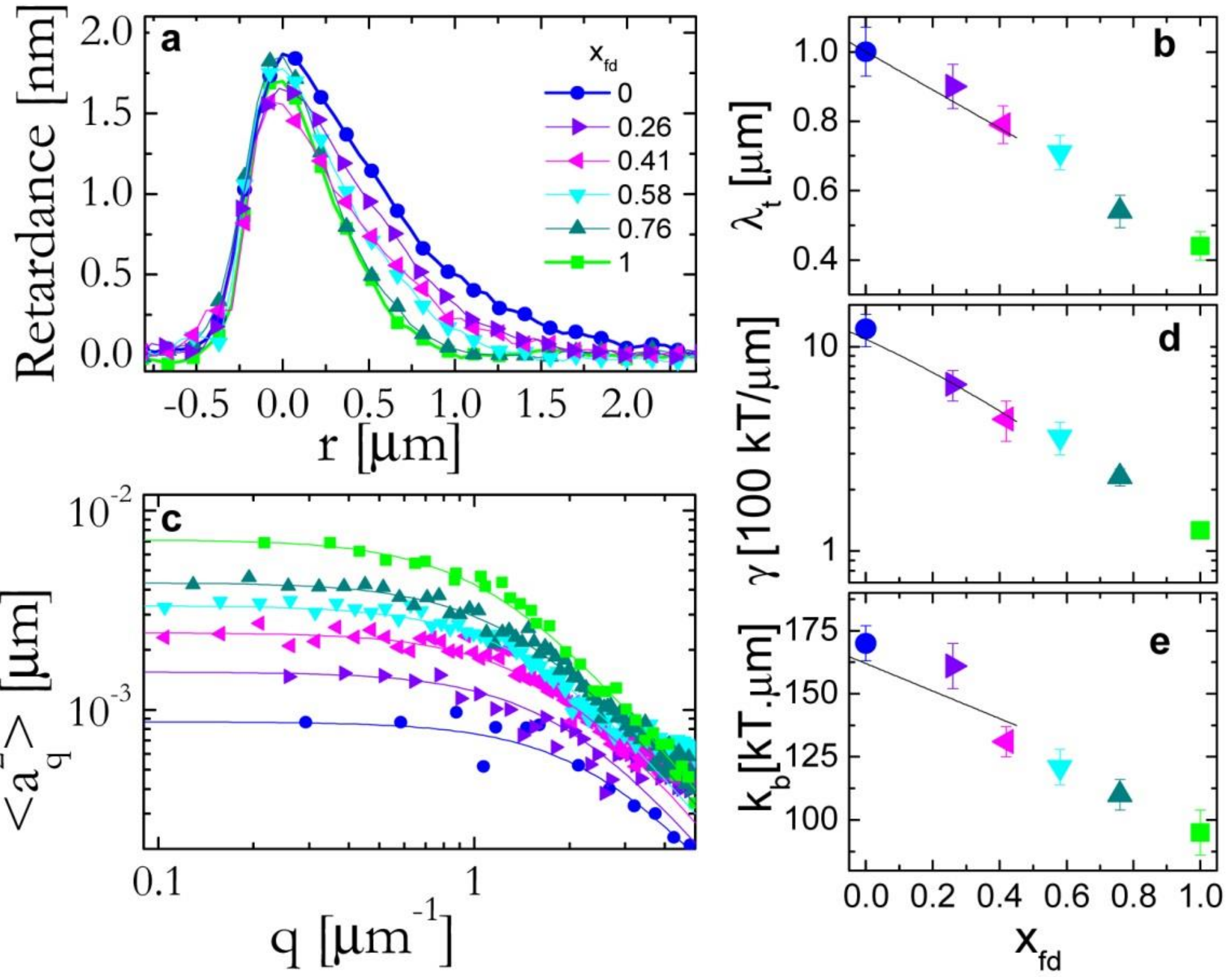

**Fig. S1: Evolution of the structure and the energetic of the edge of a colloidal membranes as a function of $x_{fd}$ at Dext=40mg/mL. a**) Retardance profile of the edge. **b**) The twist penetration length $\lambda_t$, decreases linearly as a function of $x_{fd}$. **c**) Fluctuation sprectrum of the edge of a colloidal membrane. All spectrum show similar features and can be modeled by $<a_q^2>=k_BT/(\gamma+\kappa*q^2)$, which fits yield values of edge line tension $\gamma$ and edge bending rigidity $\kappa$. **d**) $\gamma$ deacreases exponentially with $x_{fd}$. **e**) $\kappa$ decreases linearly with $x_{fd}$. Error bars come from standard deviation of 10 consecutive

measurements. Lines are fit to the data: $\lambda_t = 1 - 0.55x_{fd}$, $\gamma = 985x_{fd}^2 - 1904x_{fd} + 1088$ and $k_b = -38.5x_{fd} + 160$ in the scalloped membrane stability region ($0.04 < x_{fd} < 0.45$).

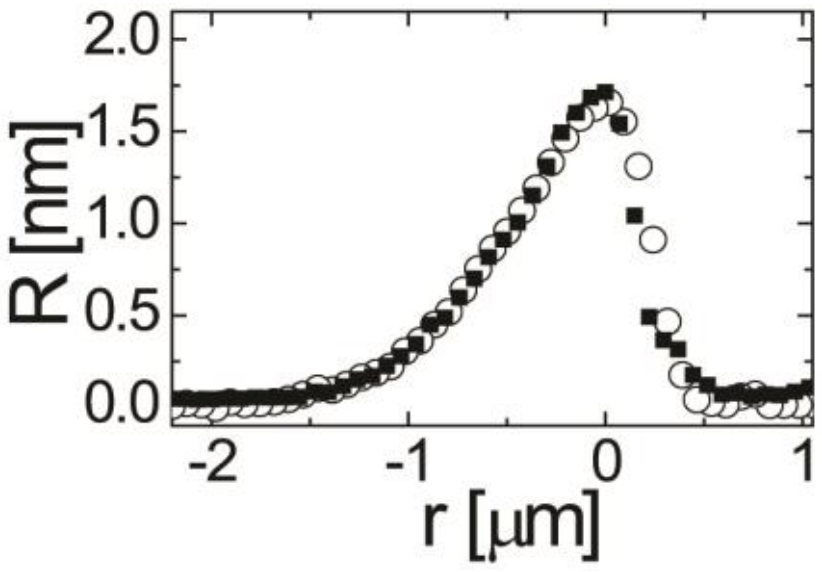

**Fig. S2: Retardance profile of the edge of the colloidal membrane between the cusp**. Left handed (circle) and right (square) handed edge display the same twist penetration length for membranes at Dext=40mg/mL, $x_{fd}$=0.26.

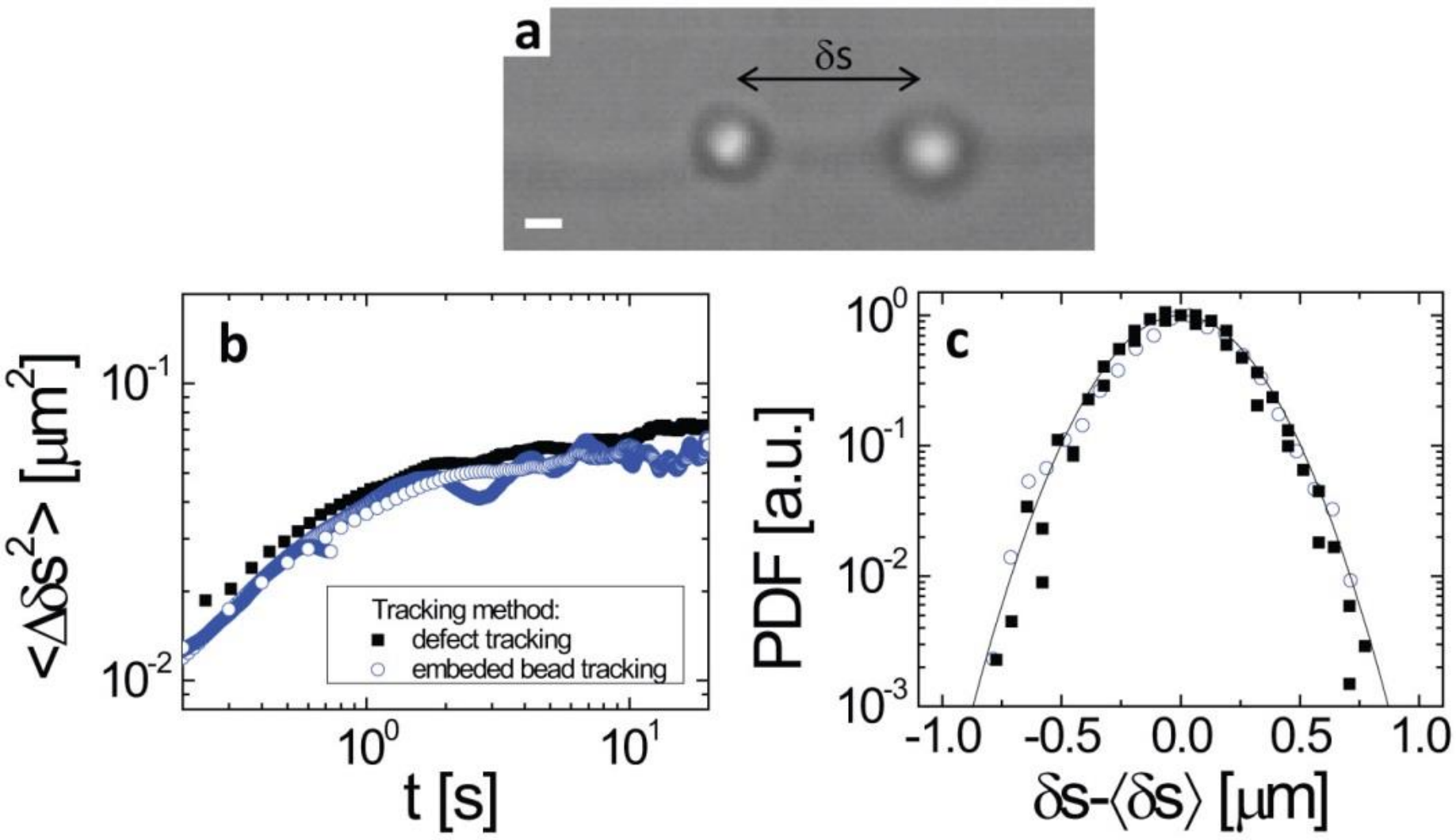

**Fig. S3: Embedding colloidal beads into the cusps**. **a**) Using laser tweezers, we can embed 1.5um diameter silica beads into the cusps. Scale bar is 1.0um. **b, c**) The mean square displacement and the PDF of the fluctuations of paired cusps with and without beads are identical. The beads do not affect the defects interactions.

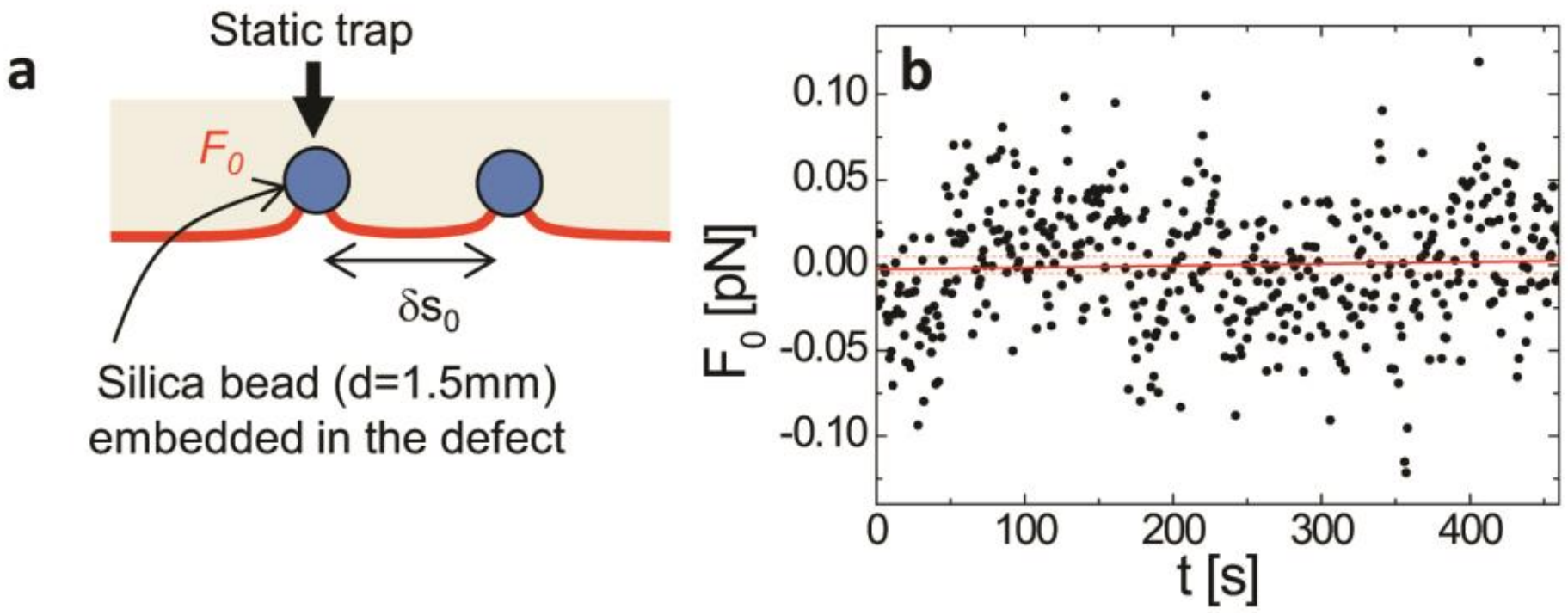

**Fig. S4: Laser tweezers experiments calibration**. **a**) We embed silica bead of diameter 1.5 um in the coupled cusps. The static trap is used to measure the force exerted in this cusp while the moving trap is use to change the distance ds between the two cusp. **b**) 0-force calibration. We use only the static trap and record QPD voltage at zero force. This experiment yields the calibration for the 0-force, Fig.S4a-b. It also shows that the drift is negligible during the experiment time.

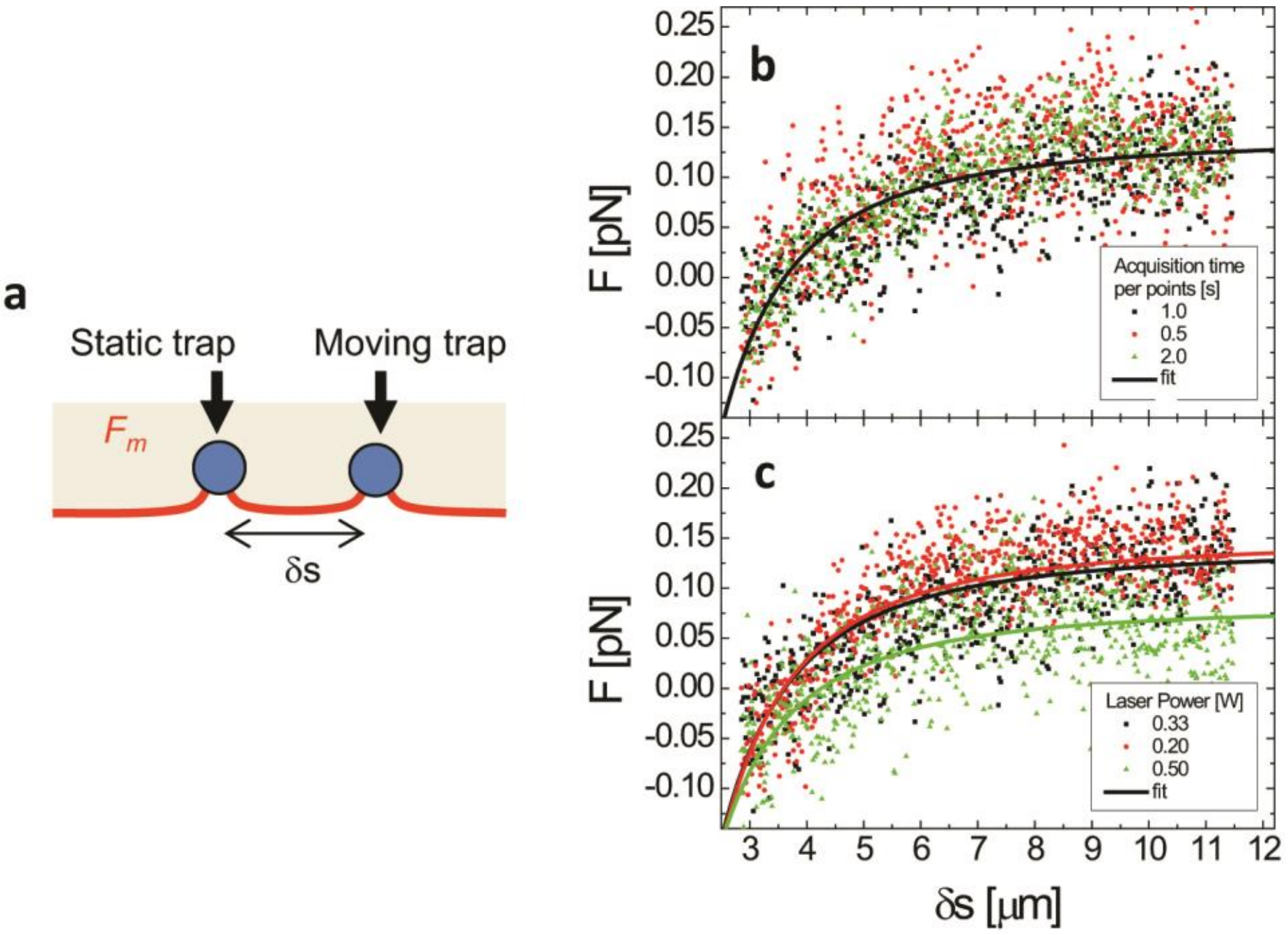

**Fig. S5: Laser tweezers experiments optimal setup**. **a**) The force is measured on bead #1 while bead #2 is moved by steps of 125nm. In between each position we record the position of bead #1 using the QPD for a time $t_D$, average the position and store it. **b**) Force measurements as a function of $t_D$. We move the trap at different speeds which allow us to determine the quasi-static regime where the measurements are independent of the speed. This is true for experiments below a speed of $t_D$=2s per step. **c**) Force measurements as a function of the trap power. The experiment is done for different trap powers to insure that the trapping does not alter the measurements. This is true for power below 0.33W

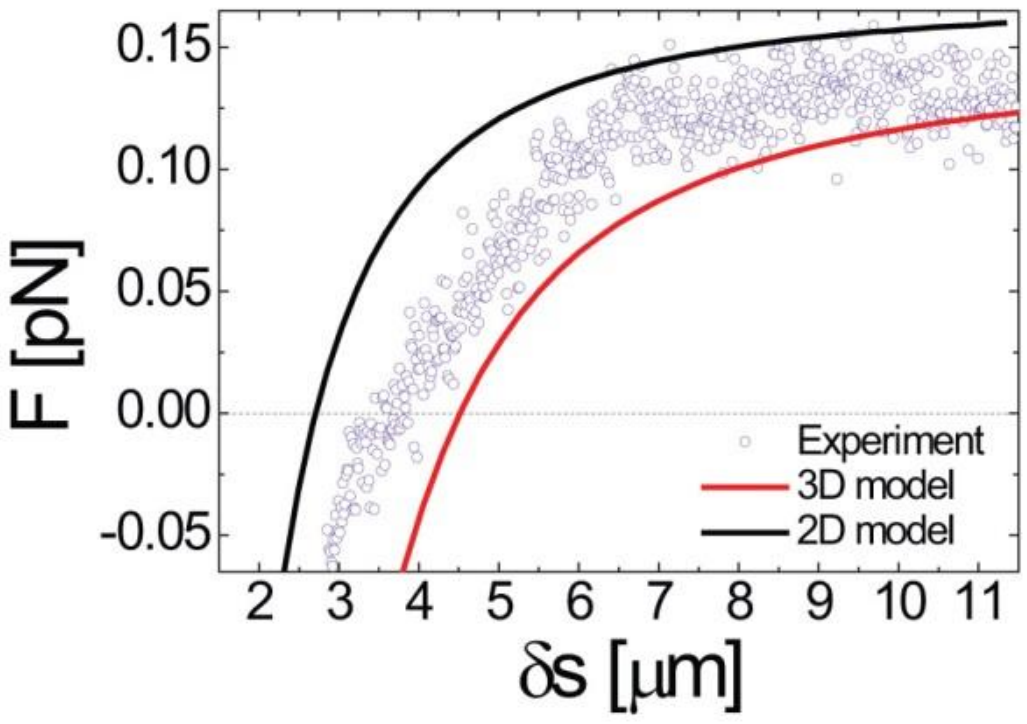

***Fig. S6: Comparison between the 2D model (no Gaussian curvature), the 3D model and the experiment at $x_{fd}$=0.06.*** *Assuming the coupled defects lie flat in the the xy plane ($A_z$=0 – no Gaussian curvature), the 2D model (black line) underestimate the equilibrium distance between coupled defects and overestimate $\alpha$ whereas the 3D model (red line) due to the Gaussian curvature includes an extra repulsive term which allows to better fit the experiments*

## Supplementary Movies

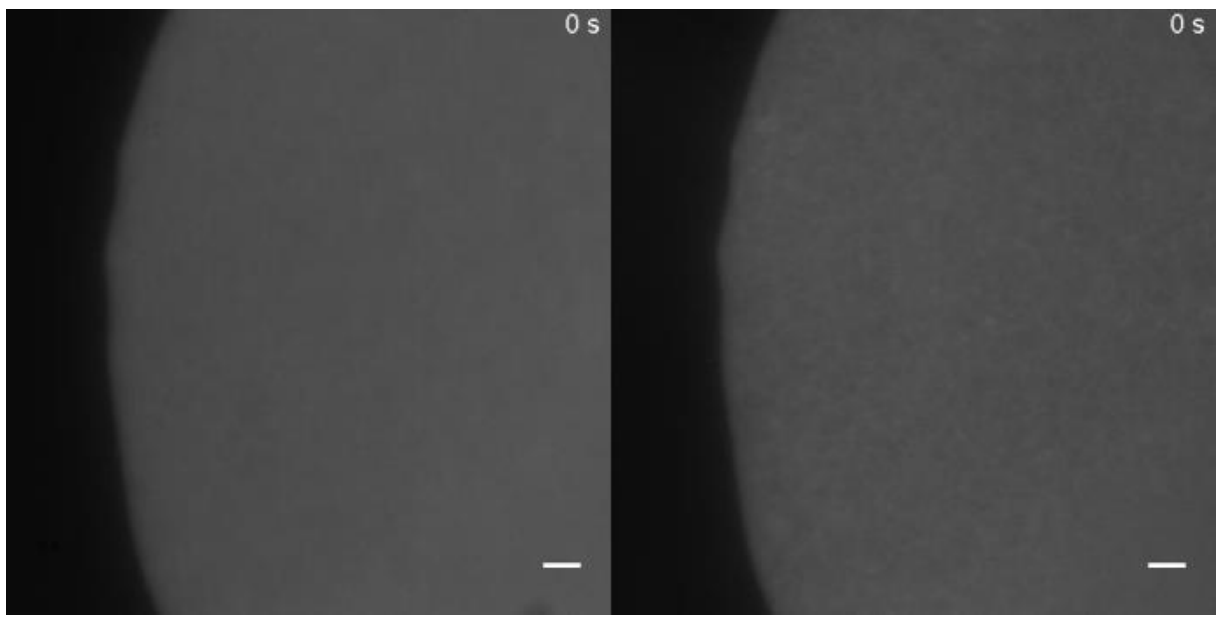

**Movie 1**: Dual view fluorescence visualization of a colloidal membrane at $x_{fd}$ =0.6 and dext=40mg/mL. The left panel correspond to the visualization of *fd* viruses labeled with alexa 488 while the left panel correspond to the visualization of the *fd*-Y21M viruses labeled with dylight 550. Scale bar is 2um.

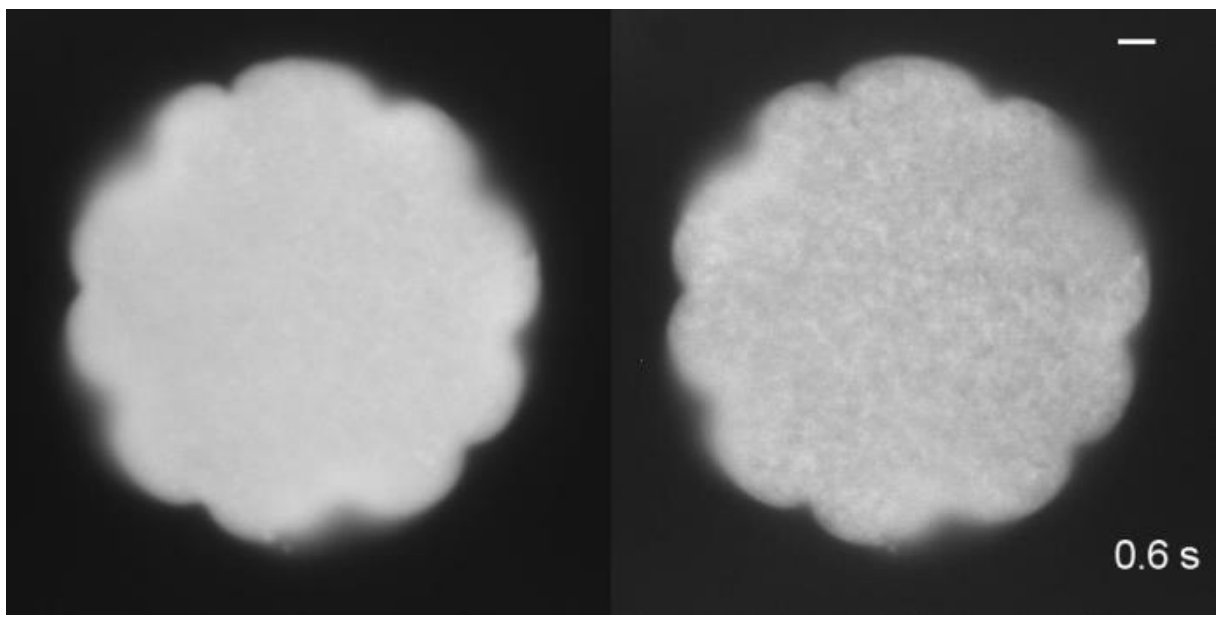

**Movie 2**: Dual view fluorescence visualization of a colloidal membrane at $x_{fd}$ =0.26 and dext=40mg/mL. The left panel correspond to the visualization of *fd* viruses labeled with alexa 488 while the left panel correspond to the visualization of the *fd*-Y21M viruses labeled with dylight 550. Scale bar is 2um

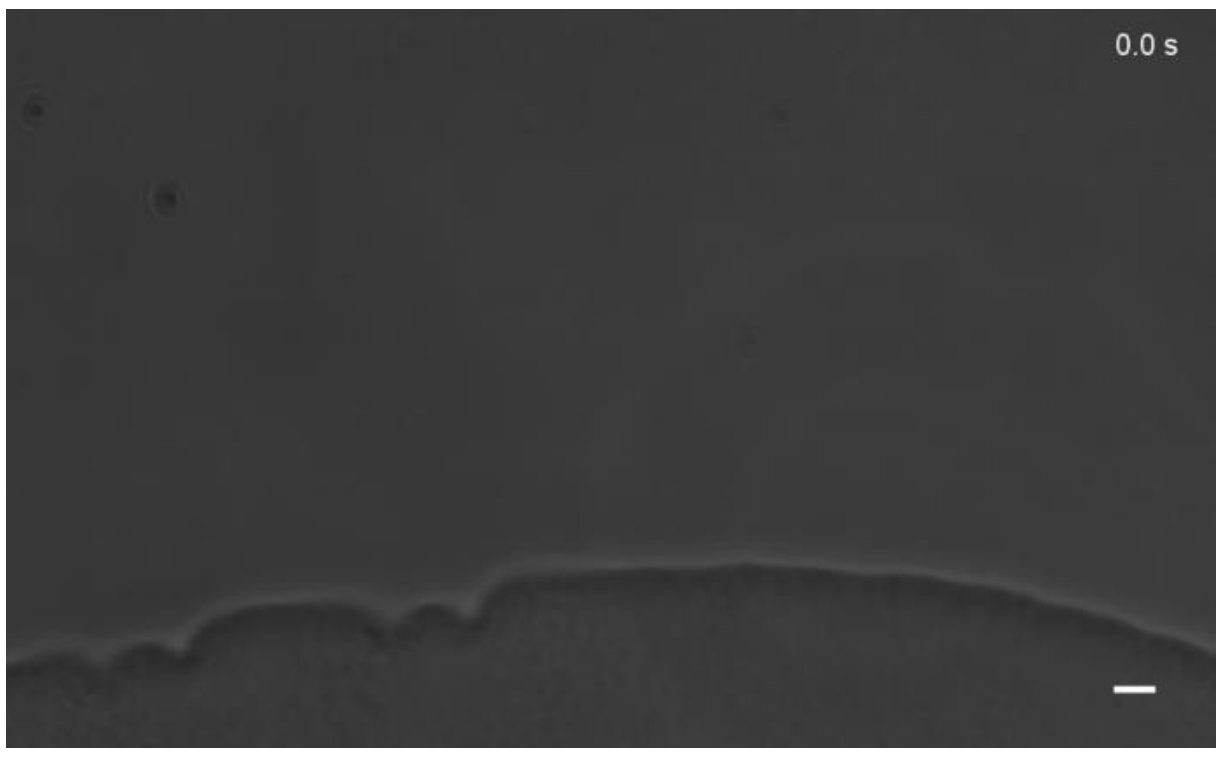

**Movie 3:** Formation of a pair of defects with laser tweezers at $x_{fd}$ =0.05 and dext=40mg/mL. Using a point trap at a power of 0.5 watt we embed a pair of defects on the edge of a colloidal membrane. The visualization is done using phase contrast microscopy. Scale bar is 2um

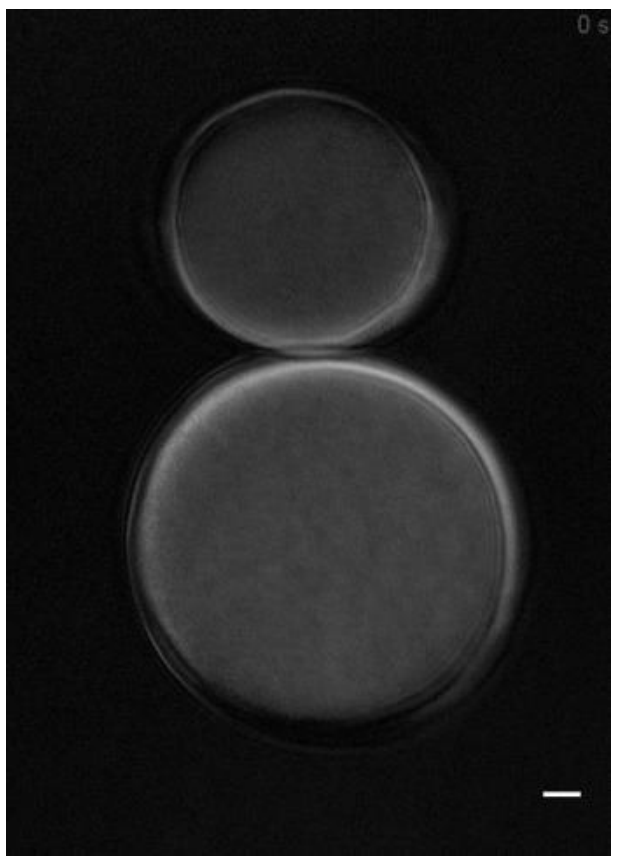

**Movie 4**: coalescence of two membrane with similar chirality at $x_{fd}$ =0.26 and dext=40mg/mL. Using angled light 2D-LC-PolScope allows us to distinguish left handed membrane from right handed membranes. Left handed membrane display a dark upper edge while right handed membranes display bright upper edge in our angled light 2D-LC-PolScope. While merging the two membranes create twisted bridges that allow the twist at the edge of the membrane to relax. Scale bar is 2um.

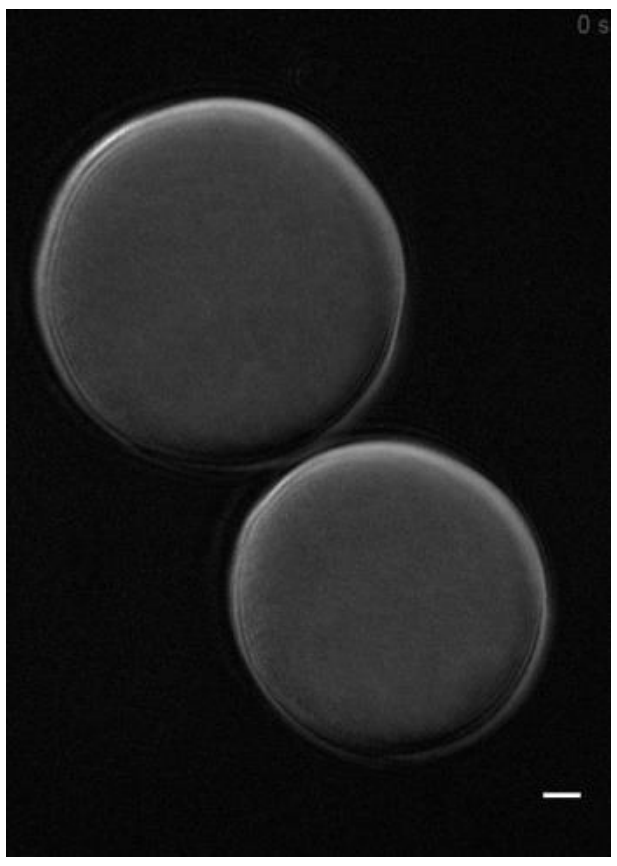

**Movie 5**: coalescence of two membrane with opposite chirality at $x_{fd}$ =0.26 and dext=40mg/mL. Using angled light 2D-LC-PolScope allows us to distinguish left handed membrane from right handed membranes. Left handed membrane display a dark upper edge while right handed membranes display bright upper edge in our angled light 2D-LC-PolScope. While merging the two membranes create a pair of point defects that separate the right handed edge from the left handed edge. Scale bar is 2um.

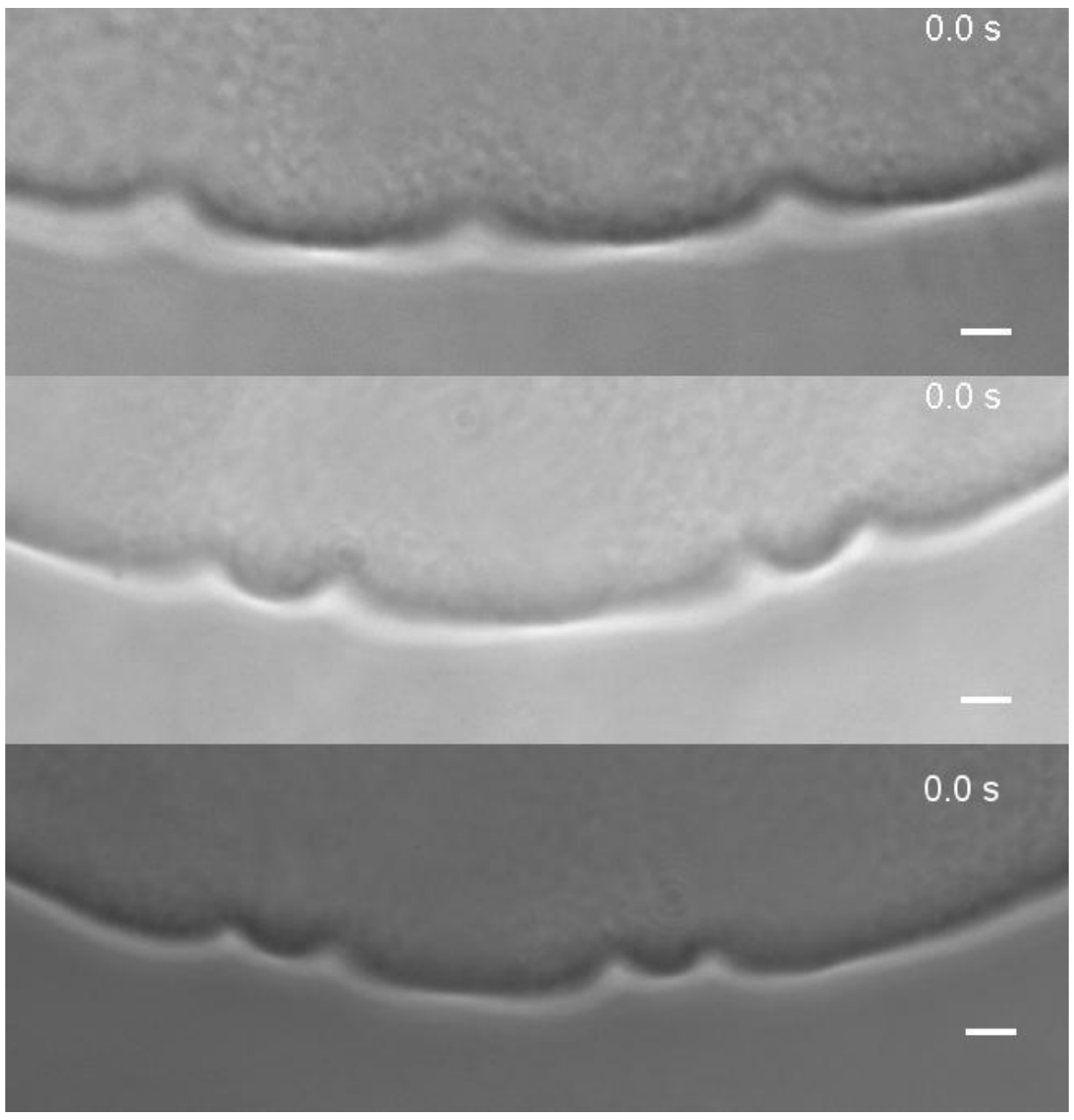

**Movie 6**: Fluctuations of the defects at dext =40mg/mL as a function of the strength of the chirality: upper panel, $x_{fd}$ =0.26; middle panel, $x_{fd}$=0.15; and lower panel, $x_{fd}$ =0.05. As $x_{fd}$ varies from the achiral limit $x_{fd}$ =0.26 to the chiral limit the defect pair together with an increasing strength. The visualization is done using phase contrast microscopy. Scale bar is 2um